\documentclass[fleqn,twoside]{article}
\usepackage{espcrc2}

\usepackage{graphicx}
\usepackage[figuresright]{rotating}

\def \SAIT #1 #2 #3 {{\em Mem.\ Soc.\ Astron.\ It.\/}  {\bf #1}  #2 #3}  
\def \MESS #1 #2 {{\em The Messenger\/} {\bf #1}, #2}
\def \ASTRNACH #1 #2 {{\em Astron. Nach.\/} {\bf #1}, #2}
\def \AAP #1 #2 {{\em Astron. Astrophys.\/} {\bf #1}, #2}
\def \AAL #1 #2 {{\em Astron. Astrophys. Lett.\/} {\bf #1}, L#2}
\def \AAR #1 #2 {{\em Astron. Astrophys. Rev.\/} {\bf #1}, #2}
\def \AAS #1 #2 #3  {{\em Astron. Astrophys. Suppl. Ser.\/} {\bf #1}  #2 #3}   
\def \AJ #1 #2 {{\em Astron. J.\/} {\bf #1}, #2}
\def \ANNREV #1 #2 {{\em Ann. Rev. Astron. Astrophys.\/} {\bf #1}, #2}
\def \APJ #1 #2 #3 {{\em Astophys. Journal\/} {\bf #1}  #2  #3}   
\def \APJL #1 #2 {{\em Astrophys.. J. Lett.\/} {\bf #1}, L#2}
\def \APJS #1 #2 {{\em Astrophys. J. Suppl.\/} {\bf #1}, #2}
\def \APSS #1 #2 {{\em Astrophys. Space Sci.\/} {\bf #1}, #2}
\def \ASR #1 #2 {{\em Adv. Space Res.\/} {\bf #1}, #2}
\def \BAIC #1 #2 {{\em Bull. Astron. Inst. Czechosl.\/} {\bf #1}, #2}
\def \JSQRT #1 #2 {{\em J. Quant. Spectrosc. Radiat. Transfer\/} {\bf #1},#2}
\def \MN #1 #2 {{\em Mon. Not. R. Astr. Soc.\/} {\bf #1}, #2}
\def \MEM #1 #2 {{\em Mem. R. Astr. Soc.\/} {\bf #1}, #2}
\def \PLR #1 #2 {{\em Phys. Lett. Rev.\/} {\bf #1}, #2}
\def \PASJ #1 #2 {{\em Publ. Astron. Soc. Japan\/} {\bf #1}, #2}
\def \PASP #1 #2 {{\em Publ. Astr. Soc. Pacific\/} {\bf #1}, #2}
\def \NAT #1 #2 {{\em Nature\/} {\bf #1}, #2}
\newcommand{\Omegachi}{\Omega_{\chi}}

\newcommand{\bbar}{\bar{b}}   

\newcommand{\be}{\begin{equation}}
\newcommand{\ee}{\end{equation}}
\newcommand{\etal}{{\em et al.}}

\usepackage{epsfig}

\newcommand{\AmS}{{\protect\the\textfont2
  A\kern-.1667em\lower.5ex\hbox{M}\kern-.125emS}}

\title{Search for Dark Matter with GLAST}
 \author{Aldo Morselli\address[Tov]{INFN Roma2  and  University of Roma "Tor Vergata", Italy },
   Andrea Lionetto\addressmark[Tov], Alessandro Cesarini\addressmark[Tov], Francesco Fucito\addressmark[Tov]
  and Piero Ullio\address[Sissa]{SISSA, Trieste, Italy } }

\begin{document}

\begin{abstract}
The detection of exotic cosmic rays due to pair annihilation of dark matter particles in 
the Milky Way halo is a viable techniques to search for supersymmetric dark matter 
candidates. The study of the spectrum of gamma-rays, antiprotons and positrons
offers good possibilities to perform this search in a significant portion of
the Minimal Supersymmetric Standard Model  parameter space. In particular the
EGRET team  have seen a convincing signal for a strong excess of emission from the
Galactic center that has no simple explanation with standard processes. We will
review the limits achievable  with the experiment  GLAST  taking into  accounts the
LEP results and we will compare this method with the antiproton and positrons
experiments.
\vspace{1pc}
\end{abstract}

\maketitle
The EGRET team  \cite{Mayer} have seen a convincing signal for a strong
excess of emission from the galactic center, with I(E) x E$^2$ peaking at $\sim$2 GeV, and  in an error circle of 0.2 degree radius
including the position l = 0$^\circ$ and b = 0$^\circ$. 
 
\begin{figure}
\centerline{\epsfig{file=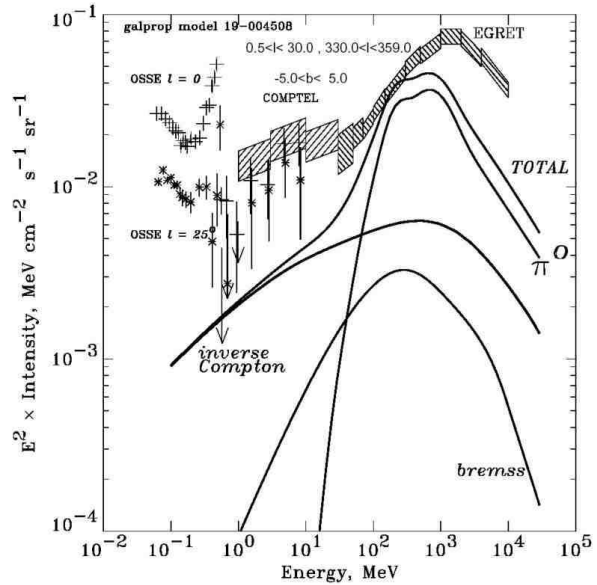,width=0.52\textwidth,angle=0,clip=}}
\caption{\label{EgretMap} \it Gamma-ray energy spectrum of the inner galaxy ($ 300^\circ \ge l \le 30^\circ $) compared with what is
expected for standard propagation models  }
\end{figure}

\begin{figure}
\centerline{\epsfig{file=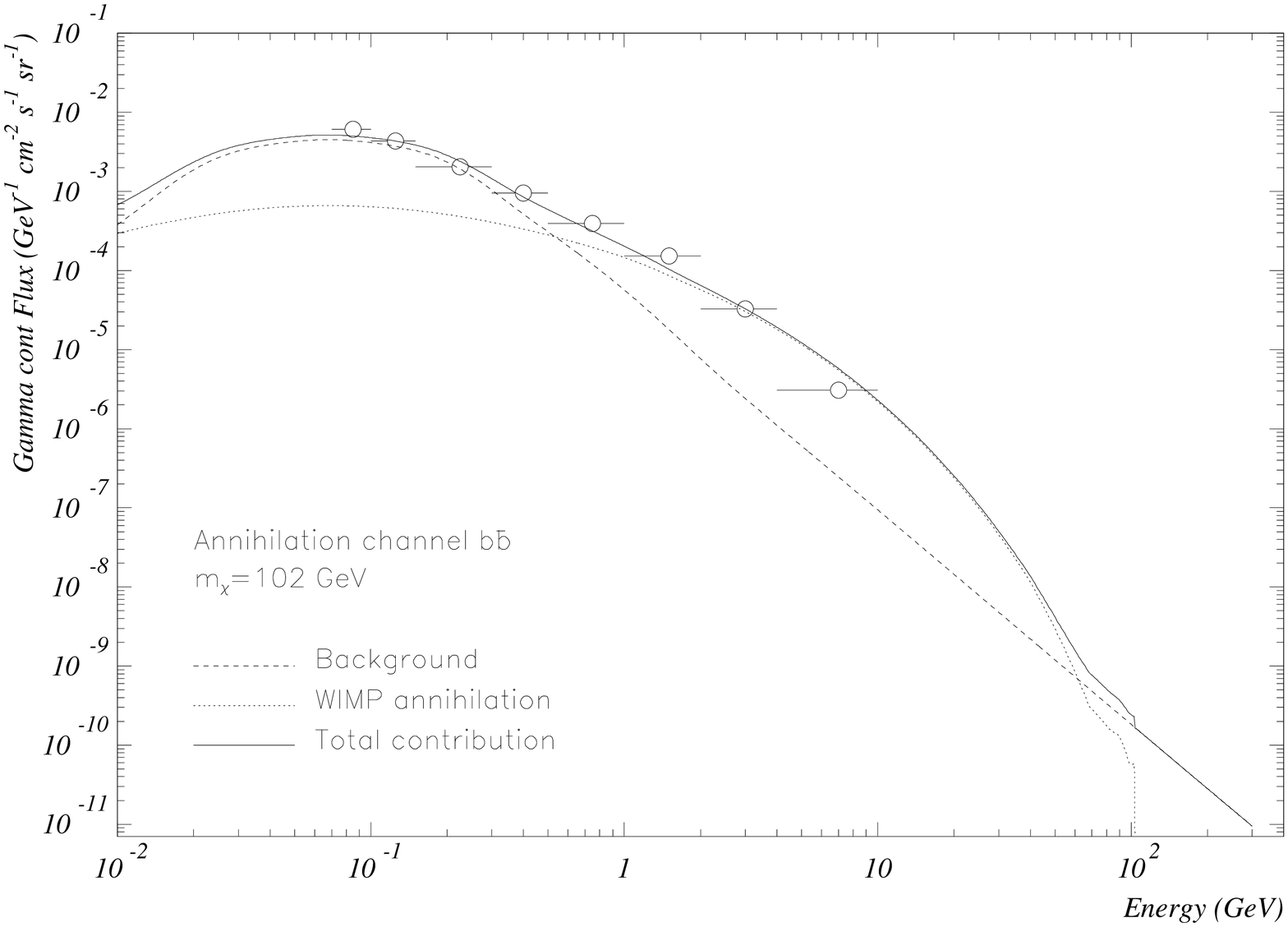,width=0.55\textwidth,angle=0,clip=}}
\caption{\label{Egretflux1} \it Gamma-ray energy spectrum of the inner galaxy ($ 300^\circ \ge l \le 30^\circ $) compared with what is
expected for standard propagation models  }
\end{figure}

This is a particular aspect of a more general problem 
 of the diffuse Galactic gamma-ray emission \cite{Hunter} that also outside the galactic center reveals  a spectrum which is 
harder than expected. As  can be seen in  figure~\ref{EgretMap}, the
spectrum observed with EGRET below 1 GeV is in agreement with the assumption that the cosmic ray spectra
and the electron-to-proton ratio observed locally are uniform,
however, the spectrum above 1 GeV, where the emission
is supposedly dominated by $\pi^\circ$-decay, is harder than that
derived from the local cosmic ray proton spectrum \cite{Strong}. 
Many different approaches are trying to solve this problem; one option is the relaxation of the 
assumption that the local cosmic ray electron spectrum is not  representative for the entire 
Galaxy and it is on average harder than that measured locally. Another possibility is that there is some variability in
the spectral indices of SNR cosmic ray sources  (for a discussion see \cite{Pohl}). 
Here we will connect the problem of the GeV excess with the problem of the missing dark matter in the Universe and we will examine the
possibility to disentangle this effect with the future space $\gamma$-ray and  cosmic ray experiments.

Over the last   years our knowledge 
of the matter and energy content of the universe has improved 
dramatically.  Astrophysical measurements from several experiments 
are now converging and a standard cosmological model is emerging.
The most significant new data come from recent 
measurements of the cosmic microwave background radiation (CMBR)\cite{debernardis}
and measurements of the Hubble flow using distant supernovae\cite{supern}.

 Current data favor (see for example \cite{astr_dark2}) a
flat universe with a cosmological constant $\Omega_{\Lambda} = 1 - \Omega_m $ and a total matter density of about
40\%$\pm$10\% of the critical density of the Universe, with a contribution of the baryonic dark matter  less then
5\%.  The remaining matter should be 
 composed of some yet-undiscovered matter form, such as Weakly Interacting Massive Particles (WIMP);
a good candidate for WIMP's 
is the Lightest Supersymmetric Particle (LSP) in R-parity conserving supersymmetric  models.  

The motivation for supersymmetry at an accessible energy is provided by the gauge hierarchy problem \cite{hierar},
namely that of understanding why $m_W \ll m_P$, the only candidate for a fundamental mass scale in physics. 
This difference introduces problems because one must fine-tune the bare mass parameter so that is almost exactly
cancelled by the quantum correction in order to obtain a small physical value of $m_W$. This seems unnatural, and
the alternative is to introduce new physics at the TeV scale and to postulate approximate
supersymmetry\cite{super}, where boson and fermion partners naturally induce cancellations in  quantum corrections
in case 
$$ | m_B^2 -  m_F^2 | \le 1TeV $$.

This is also the reason to expect that, if supersymmetry is real, it might be accessible to the current generation
of accelerators and in the range expected for a cold dark matter particle.

The minimal supersymmetric extension of the Standard Model (MSSM) \cite{mssm} has the same gauge interactions as
the Standard Model; most often its phenomenology is simplified by reducing its parameter space to only six parameters:
the higgs mixing parameters $\mu$  that appears in the neutralino and chargino mass matrices, 
the common mass for scalar fermions at the GUT scale $m_0$, the gaugino mass parameter $M_{1/2}$, the trilinear 
scalar coupling parameter A, the ratio
between the two vacuum expectation values of the Higgs fields defined as $\tan \beta = v_2 /v_1 = <H_2> / <H_1>$ 
and the mass of the pseudoscalar Higgs $m_A$.

The LSP is expected to be stable in R-parity conserving versions of the MSSM, and hence should be present in the Universe today as a cosmological
relic from the Big Bang \cite{relic}. R-parity is a discrete symmetry related to baryon number, lepton number and spin:
$$ R=(-1)^{3B+L+2S}$$
It is easy to check that R=+1 for all Standard Model particles and R=-1 for all their supersymmetric partners.
There are three important consequences of R conservation: (i) sparticle are always produces in pairs; (ii) heavier
sparticles decay into lighter sparticles and (iii) the LSP is stable because it has no allowed decay mode.

The LSP is expected also to be  neutral, because with an electric charge or strong interaction,
it would have condensed along with ordinary baryonic  matter during the formation of astrophysical structures,
and should be present in the Universe today in anomalous heavy isotopes \cite{isotop}.
This leaves as candidates a sneutrino with spin 0, the gravitino with spin 3/2  and the neutralino
$\chi$ that is a combination of the partners of the $\gamma$, $Z$ and the neutral Higgs particles (spin 1/2).

Searches for the interactions of relic particles with nuclei rule out sneutrinos with masses lighter than
few TeV \cite{sne}, while the gravitino could constitute warm dark matter 
with a mass around 1 keV. So the best candidate for cold dark matter appears to be the neutralino $\chi$.
The experimental LEP lower limit on $m_{\chi}$ is 
\cite{masslim} $$m_{\chi} \ge 50 ~ GeV $$  
As $m_{\chi} $ increases, the LSP pair annihilation rate in dark structures decreases, but, as we will show below,  
up to $\sim 300 ~ GeV$,  the distortion of the
spectrum  of the diffuse gamma ray background due to a neutralino induced component up to
the neutralino mass can be a possible signature of the existence of the LSP\cite{dark2}. 

How can this kind of signal be seen? 

In figure~\ref{Egretflux1}  is shown the EGRET data from the Galactic center, the diffuse gamma ray background flux expected from the standard interactions and propagation models of cosmic ray protons and electrons and an example of the flux due to neutralino annihilation in the dark matter halo. In this case the signal is for a $\sim$ 100 GeV neutralino
and for the $b \bbar$ annihilation channel (the spectral shape of the other channels is 
very similar).

The total flux is:
\begin{equation}
\phi_\gamma=N_b\phi_b+N_\chi \phi_\chi,
\label{totflux}
\end{equation}
where
$N_b$ and $N_\chi$ are normalizations parameters for the standard and exotic flux, respectively.
 $N_{b}$  is associated to the interstellar medium column density $n(l)$, integrated along the line of sight.

The background flux can be written in the following way:
\begin{equation}
\phi_{b}(E_\gamma)=Em(E_\gamma)/\left( cm^2 sr\right)
\end{equation}
\begin{equation}
N_b=\int_{l.o.s}\frac{dl}{4\pi}n(l)/\left( cm^{-2}sr^{-1}\right)
\label{nback}
\end{equation}
where $Em(E_\gamma)$ [$GeV^{-1}s^{-1}$] is the emissivity per hydrogen atom, which gives the number of secondary photons with energy $E_\gamma$, emitted per unit time and unit energy.
The two main production mechanisms for the background are through the $\pi^{0}$ production and the inverse compton scattering.  We have considered two channels for the $\pi^{0}$ production, i.e. primary $p$ and $He$. The processes are of the type:
\begin{equation}
p+X\to..\to\pi^{0}\to2\gamma,  
\end{equation}
$$ ~~~~~~He+X\to..\to\pi^{0}\to2\gamma$$
where $X$ can be interstellar $H$ or $He$ nucleus.
We have simulated the spectrum using the Galprop computer code \cite{smapj}. The WIMP flux  $\phi_\chi$ in eq.(\ref{totflux}) originates from the annihililation of a couple of generic WIMPs, of mass $m_\chi$, in allowed tree-level
final states, of which the leading channels are usually $b\bar{b},c\bar{c},t\bar{t},W^+W^-,Z^0Z^0$. Light quark states are suppressed by $m_{f}^2/m_{\chi}^2$ in the annihilation cross section, where $m_{f}$ is the intermediate fermion mass. Higgs bosons annihilation states are included, allowing for their decay into particles for which we do simulate. The continuum gamma ray flux coming from a direction that forms an angle $\psi$ with the direction of the Galactic center can be written as:
\begin{equation} \label{gammafluxcont}
\phi(E,\psi)=\frac{2\sigma v}{4\pi m_\chi^2}\int_{l.o.s} \rho^2 (l)dl(\psi)
\end{equation}
where $m_\chi$ is the WIMP mass, $\sigma v$ is the total annihilation cross section times the relative velocity of the two annihilating WIMPs (in the galactic halo $v/c\sim10^{-3}$) and $\rho(l)$ is the WIMP mass density along the line of sight.  Contributions along the line of sight are then summed. The equation (\ref{gammafluxcont}) can be factorized into two pieces, one depending only on the cross section and the WIMP mass, and the other one depending on the WIMP distribution in the galactic halo. This factor depends in a crucial way from the halo model used.
We now write the differential flux $\phi_\chi$ of eq.(\ref{totflux}), in units of $cm^{-2} s^{-1} sr^{-1} GeV^{-1}$, as:
\begin{eqnarray}
\phi_\chi (E,\psi) &=&1.87 \cdot 10^{-11}\left( \frac{2\sigma v}{10^{-29} cm^3 s^-1}\right) \times
\nonumber \\
&& \times  \left( \frac{10 GeV}{m_\chi}\right)^2   \cdot \left( \frac{dN}{dE}\right) \label{wimpflux}
\end{eqnarray}
where dN/dE is the the number of photons produced per unit energy in each intermediate annihilation channel. We then introduced the dimensionless function $J(\psi)$ that encodes the dependence from the halo density profile:
\begin{equation}
J(\psi)=\frac{1}{8.5 Kpc} \left(\frac{1}{0.3 GeV/cm^{3}}\right)\int  \rho^2 (l)dl(\psi)
\end{equation}
In such a way the adimensional normalization factor $N_\chi$ of eq.(\ref{totflux}) is then precisely $J(\psi)$. Actually we have averaged the value of $J(\psi)$ over a solid angle $\Delta\Omega$ around the direction determined by the angle $\psi$:
\begin{equation}
\left< J(\psi)\right>_{\Delta\Omega}=\frac{1}{\Delta\Omega}\int J(\psi)d\Omega
\label{jpsiave}
\end{equation}
where we have chosen the same EGRET region around the galactic center of $\Delta\Omega=2.15\cdot 10^{-3}\: sr$.

As it can be seen from figure~\ref{Egretflux1}, the fit to the data greatly improve 
when a neutralino component  is added. Of course this cannot be assumed as evidence for the existence of supersymmetry particles as the dark matter component of the halo, but as an indication that more experiments with greater sensitivity and exposure are needed.

In the next session we present one future possibility, i.e. the experiment GLAST.

\begin{figure}[ht]
 \centerline{\epsfig{file=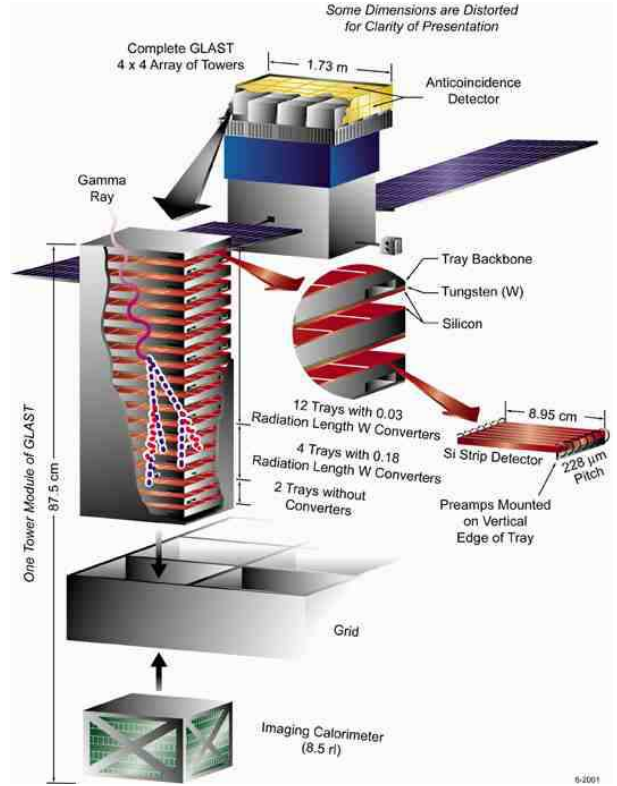,width=0.5\textwidth,angle=0,clip=}}
\vspace{-0.5cm}
\caption{\label{glastscheme} \it The GLAST instrument, exploded to show
the detector layers in a tower, the stacking of the CsI
logs in the calorimeter, and the integration of the subsystems.}
\end{figure}

\section{The Gamma-ray Large Area Telescope GLAST}	

The standard techniques for the detection of gamma-rays in the pair production regime energy range  are very different from the X-ray
detection.

 \begin{figure}
\vspace{0.5cm}
    \centerline{\epsfig{file=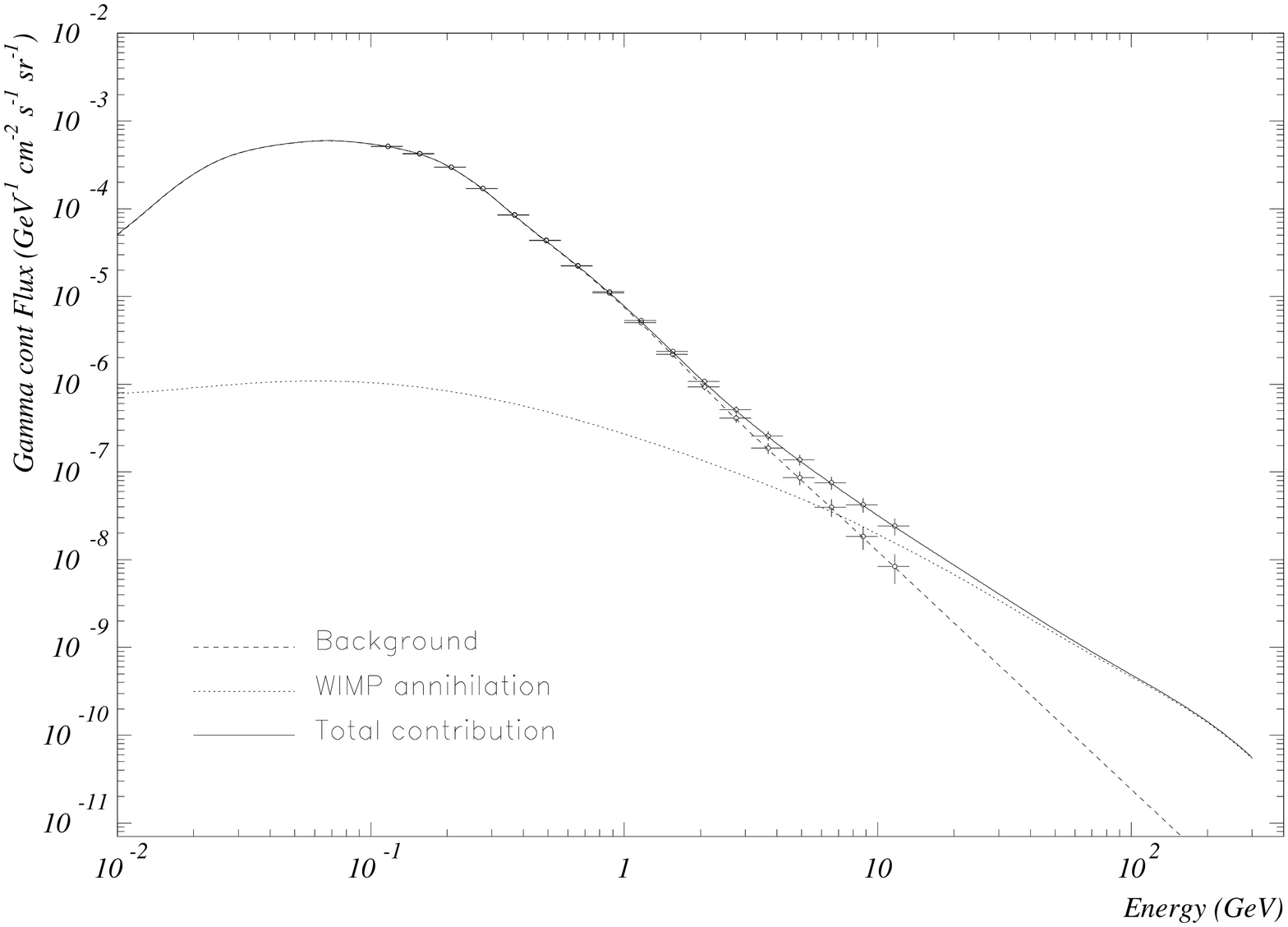,width=0.55\textwidth,angle=0,clip=}}
\vspace{-0.7cm}
\caption{\label{gline2} \it Total photon spectrum from the
galactic center from standard propagation models and from one neutralino annihilation models
and the kind of statistical errors that it is expected in three years with GLAST.  }
\end{figure}

\begin{figure}
    \centerline{\epsfig{file=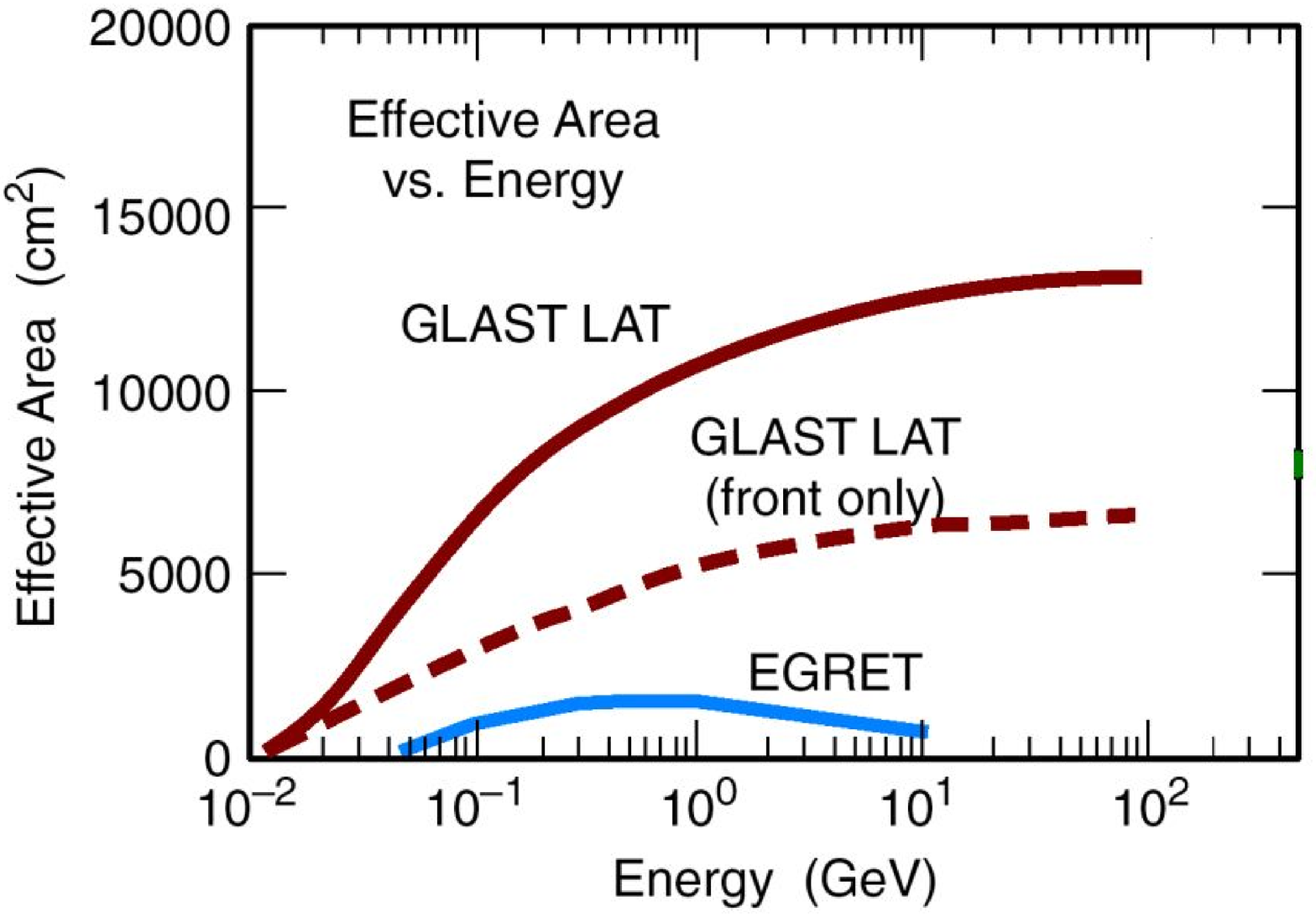,width=0.45\textwidth,angle=0,clip=}}
\vspace{-0.6cm}
\caption{\label{aff} \it GLAST  effective area  as a function of energy including  all background and track quality cuts compared with EGRET's one}
\vspace{0.5cm}
    \centerline{\epsfig{file=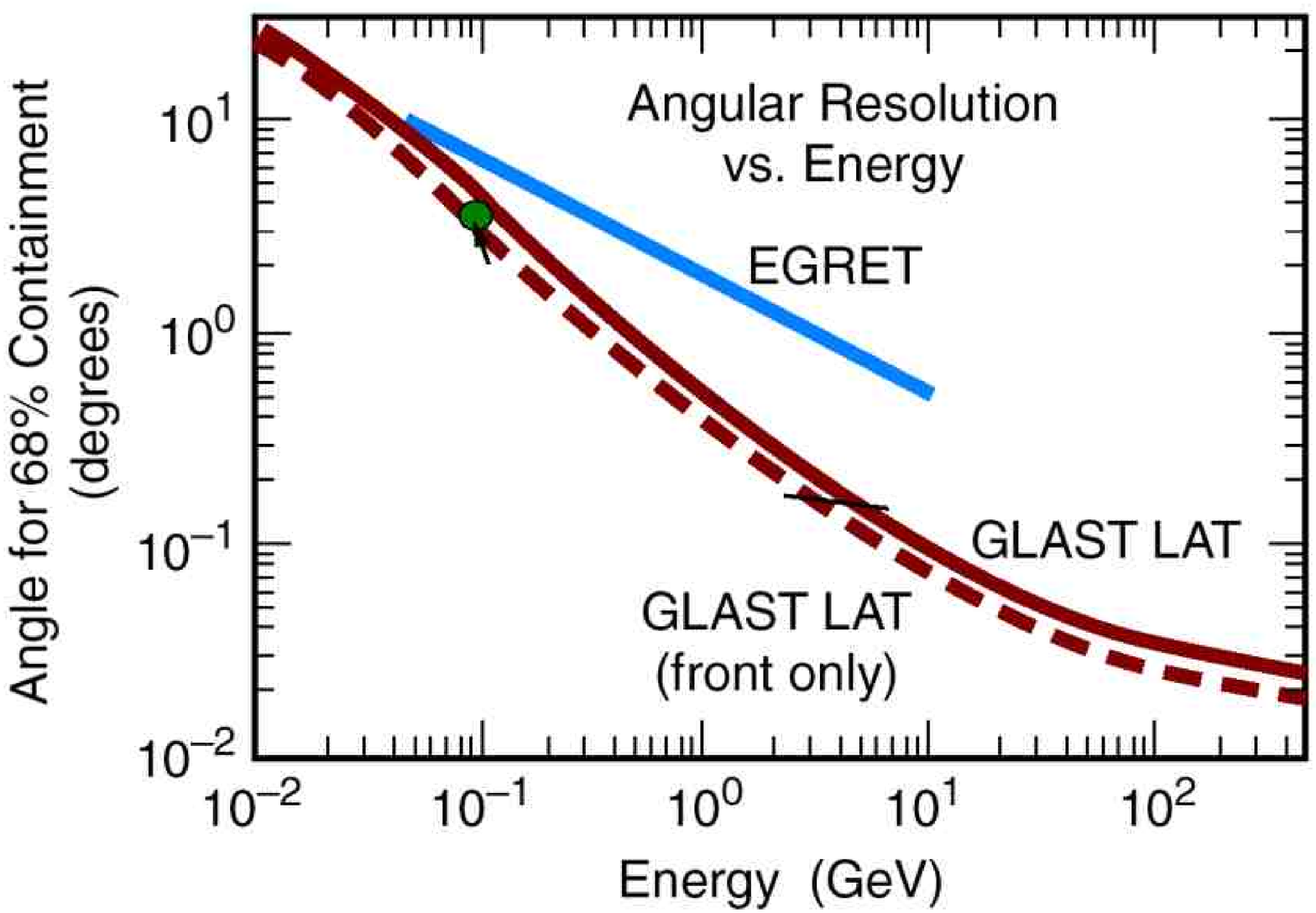,width=0.45\textwidth,angle=0,clip=}}
\vspace{-0.7cm}
\caption{\label{ang} \it GLAST  angular resolution.}
\vspace{0.5cm}
    \centerline{\epsfig{file=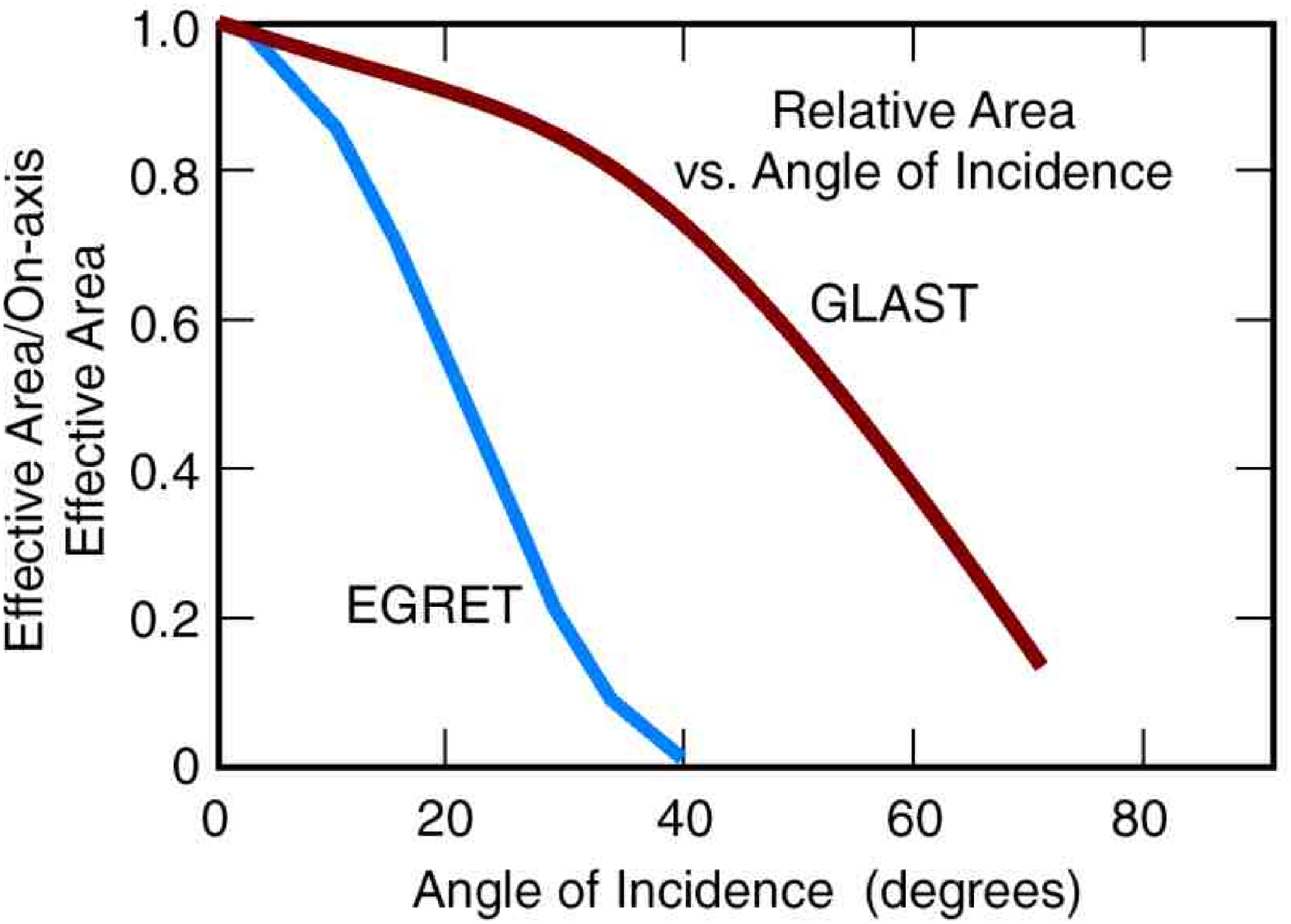,width=0.45\textwidth,angle=0,clip=}}
\vspace{-0.6cm}
\caption{\label{relat} \it GLAST  relative effective area for
1 GeV photons as a function of the zenith angle}
\end{figure}

\begin{figure}
\vspace{0.5cm}
    \centerline{\epsfig{file=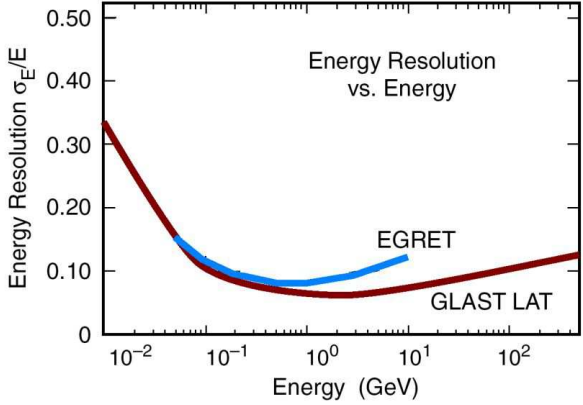,width=0.5\textwidth,angle=0,clip=}}
\vspace{-0.6cm}
\caption{\label{enres} \it GLAST and EGRET energy resolution }
\end{figure}

For X-rays detection focusing is possible and this permits large effective area, excellent energy resolution, very low background.
For gamma-rays  no focusing is  possible and this means limited effective area,  moderate energy resolution, high background but a wide field
of view. This possibility to have a wide field of view is enhanced now, in respect to EGRET, with the use of silicon detectors, that allow 
a further increase of the ratio between height and width, essentially for two reasons: a) an increase of
the  position resolution that allow a decrease of the distance between the planes of the tracker without affecting the angular resolution,  b)
the  possibility to use the silicon detectors themselves for the trigger of an events, with the elimination of the Time of Flight system, that
require some height. 
The Gamma-ray Large Area Space Telescope (GLAST)\cite{glast},  has
been selected by NASA  as a  mission involving an international collaboration of  particle physics and astrophysics 
communities from
the United States, Italy, Japan, France and Germany  for a launch in the first half of 2006. 
The main scientific objects are
the study of all gamma ray sources such as blazars, gamma-ray bursts,  supernova
remnants, pulsars, diffuse radiation, and unidentified high-energy sources.
Many years of refinement has led to the configuration of the apparatus shown in figure~\ref{glastscheme},
where one can see the  4x4 array of identical towers each formed by:
$\bullet $   Si-strip Tracker Detectors and converters arranged in 18 XY tracking planes for the measurement
of the photon direction.
$\bullet $ Segmented array of CsI(Tl) crystals for the measurement the photon energy.
$\bullet $ Segmented Anticoincidence  Detector (ACD).
The main characteristics, shown in figures~\ref{aff} to ~\ref{enres},
are an  energy range between	20 MeV and 300 GeV, 
a field of view of $\sim$	3  sr,  an energy resolution	of $\sim$ 5\% at 1 GeV, 
a point source sensitivity of  2x10$^{-9}$ (ph~cm$^{-2}$~s$^{-1}$)     at  0.1 GeV,          
an event deadtime	 of 20 $\mu s$ and a peak effective area  of  10000 cm$^2$,  
for a required power	of  600 W and a payload weight of  3000 Kg.

 The list of the
people and  the Institution involved in the collaboration together with the on-line
status of the project is available at {\sl http://www-glast.stanford.edu}. 
A description of the apparatus can be found in \cite{Bellazzini} and a description of the main 
physic items can be found in \cite{morselli}.

In figure~\ref{gline2}  is shown the total photon spectrum from the
galactic center from standard propagation models and from one neutralino annihilation models
and the kind of statistical errors that it is expected in three years with GLAST in the case of a moderate value of $J(\psi)$  ($\sim 500$) which is within the allowed ranges of both the modified isothermal and cuspy halos.
This effort will be complementary to a similar search for neutralinos looking with 
cosmic-ray experiments like the next space experiments PAMELA\cite{pam1} and AMS\cite{battiston} at the distortion of the secondary 
positron fraction and  secondary antiproton flux induced by a signal from a heavy neutralino. 
 
\begin{figure}
    \centerline{\epsfig{file=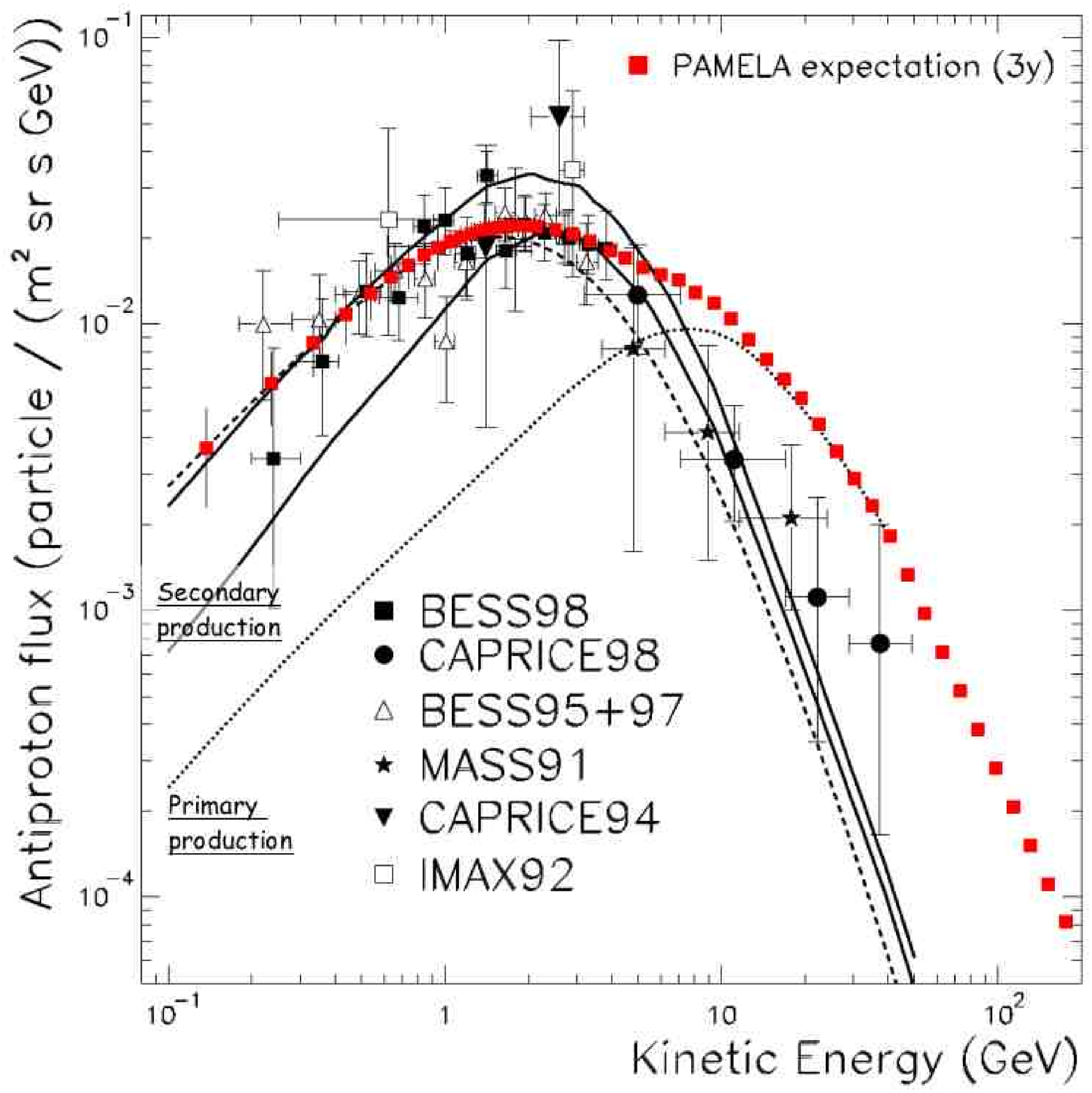,width=0.5\textwidth,angle=0,clip=}}
\vspace{-0.3cm}
\caption{\label{pamelaa} \it  Distortion of the   secondary antiproton flux induced by a signal from a heavy neutralino. The PAMELA expectation in the case of  exotic contributions are shown by black squares} 
\vspace{0.4cm}
    \centerline{\epsfig{file=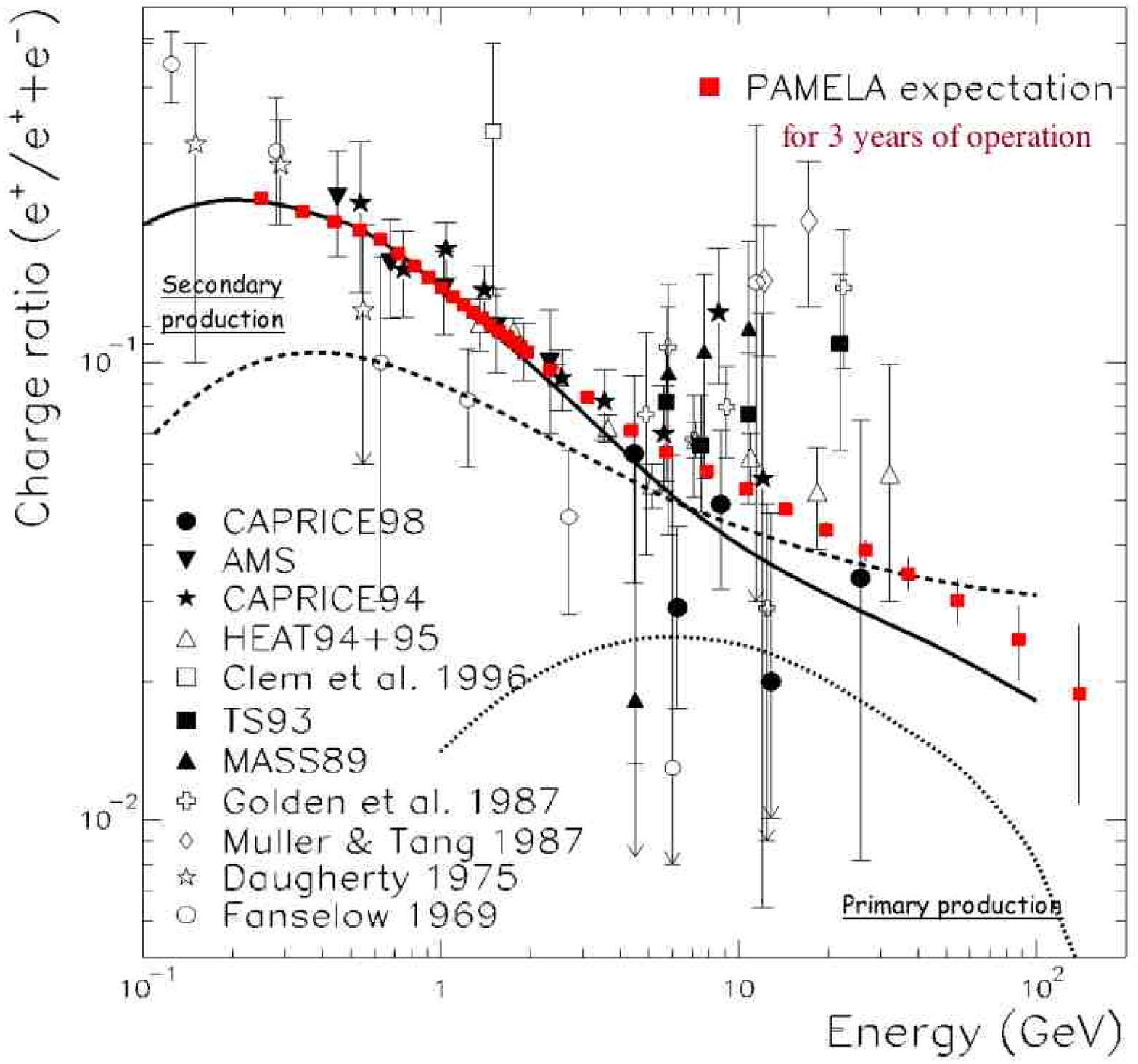,width=0.5\textwidth,angle=0,clip=}}
\vspace{-0.5cm}
\caption{\label{pamelap} \it  Distortion of the secondary 
positron fraction  The PAMELA expectation in the case of  exotic contributions are shown by black squares} 
\end{figure}

\begin{figure}[ht]
    \centerline{\epsfig{file=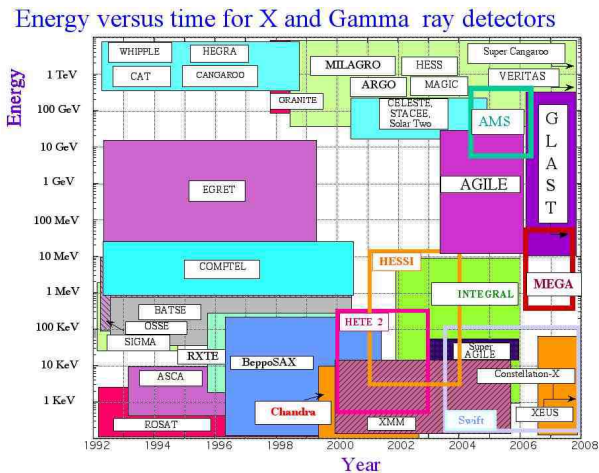,width=0.5\textwidth,angle=0,clip=}}
\caption{\label{timeline2} \it  Timeline schedule versus the energy range covered by present and future
detectors in X and gamma-ray astrophysics.  }
\end{figure}
\begin{figure}[ht]
    \centerline{\epsfig{file=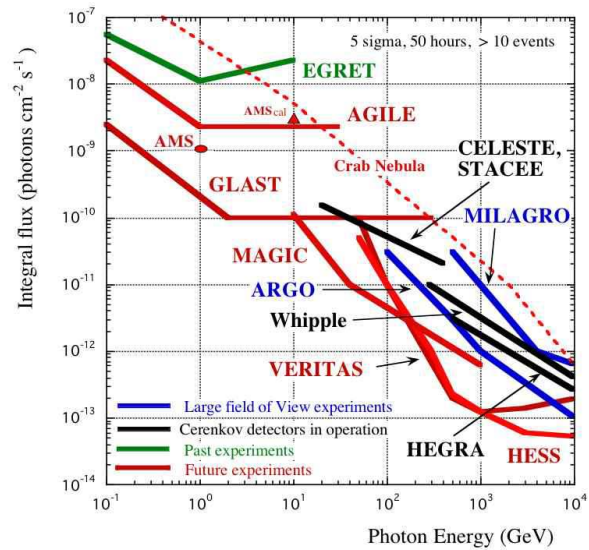,width=0.5\textwidth,angle=0,clip=}}
\caption{\label{sens2} \it Sensitivity of present and future detectors in  the
gamma-ray astrophysics.  }
\end{figure}

\section{ Conclusion}

The gamma-ray space experiment  GLAST is under construction. Its time
of operation and  energy range is shown together with the other space X-ray satellite and gamma-ray
experiments in figure~\ref{timeline2}.  Note that it will cover an interval not covered by any other
experiments. Note also the number  of other experiments in other frequencies that will allow
extensive multifrequency studies.
In the last decade, ground-based instruments have made great progress, both
in technical and scientific terms. High-energy gamma rays can be observed from the ground by experiments
that detect the air showers produced in the upper atmosphere. 
 In ~\ref{sens2}   the   GLAST sensitivity is compared with the others present and future
detectors in the gamma-ray astrophysics  range is shown.
The predicted sensitivity of a number of operational and proposed  Ground based Cherenkov telescopes,
 CELESTE, STACEE, VERITAS, Whipple is for a 50 hour exposure on a single source. EGRET, GLAST, MILAGRO, ARGO 
and AGILE sensitivity is shown for one year of all sky survey. The diffuse background assumed is
$2\cdot10^{-5} ~photons~cm^{-2} s^{-1} sr^{-1}(100 ~MeV/E)^{1.1}$, typical of the background seen by EGRET at high galactic latitudes.
The source differential photon number spectrum is assumed to have a power law index of -2, typical of many of the sources observed by EGRET. Above 1GeV
the GLAST sensitivity shown is not limited by background, but rather by the  requirement that the number of source photons detected is at least 5 sigma above the background. The AMS sensitivities is from \cite{ams_elba}. 
  Note that on 
ground only MILAGRO and ARGO will observe more than one source simultaneously.  The Home Pages of the various instruments are at
{\sl http://www-hfm.mpi-hd.mpg.de/CosmicRay/CosmicRaySites.html}.
 As is shown in \cite{rev},  indirect dark matter searches and traditional particle searches are highly complementary and, in the next few
years, many  experiments will be sensitive to
the various potential neutralino annihilation products. These include
under-ice and underwater neutrino telescopes, atmospheric Cerenkov telescopes, high altitude extensive air showers detectors  and the
 already  described space
 missions  GLAST,  PAMELA together with  AMS. In many cases,
these experiments will improve current sensitivities by several orders of
magnitude and probably, as it is shown in \cite{rev}, " all models with charginos or sleptons lighter than 300 GeV will produce observable signals in at least one
experiment in the cosmologically preferred regions of parameter space with $0.1 <
\Omegachi h^2 < 0.3$ " before LHC.

  \section*{Acknowledgements}
  We would like to thanks all the component of the GLAST Dark Matter working group for lots of discussion on the argument, 
  in particular  Elliot Bloom and Eduardo do Couto e Silva; Hans Mayer-Hasselwander for discussion on the 
  EGRET data from the galactic center;   Roberto Battiston and Franco Cervelli for the  organization of this very nice workshop and discussions.

\end{document}